\documentclass[a4paper,allowtoday]{quantumarticle}
\pdfoutput=1
\usepackage[utf8]{inputenc}
\usepackage[T1]{fontenc}
\usepackage{amsmath, amssymb}
\usepackage{graphicx}
\usepackage{hyperref}
\usepackage{doi}
\usepackage[numbers,sort&compress]{natbib}
\usepackage{physics}
\usepackage{microtype}
\usepackage{booktabs}
\usepackage[a4paper, top=2cm, bottom=2.5cm, left=3cm, right=2.5cm]{geometry}

\title{Modeling Charge Noise in Superconducting  Qubits Using Memory Multi-Fractional Brownian Motion}

\author{Mahboob Ul Haq}
\affiliation{Department of Physics, Govt Post Graduate College Timergara Affiliated with University of Malakand, Dir Lower, Pakistan}
\email{mahboobulhaqmahboob455@gmail.com}  % Replace with your email

\date{\today}

\begin{document}
	
	\maketitle
	
	\begin{abstract}
		We introduce a novel stochastic model for charge noise in superconducting charge qubits based on memory multi-fractional Brownian motion (mmfBm), capable of capturing non-stationary and long-memory effects. This framework reproduces key experimental features of decoherence and offers new insights into environmental interactions with superconducting quantum devices. 
	\end{abstract}
	
	\section{Introduction}
	
	Quantum computing has rapidly evolved from theoretical foundations \cite{feynman1982,deutsch1985,Shor1994,Grover1996} to experimental demonstrations of quantum supremacy \cite{Arute2019} and industrial-scale hardware development \cite{IBM2025,Gidney2025}. Among leading platforms, superconducting qubits have emerged as a promising building-block due to their scalability, compatibility with lithographic fabrication, and rapid gate operations. Particularly, the transmon qubit—a charge-insensitive variant of the superconducting charge qubit—offers enhanced coherence properties by optimizing the ratio of Josephson energy ($E_J$) to charging energy ($E_C$).
	
	Despite these advances, decoherence remains a critical bottleneck in realizing fault-tolerant quantum computation. Superconducting qubits are macroscopic quantum systems and hence highly susceptible to environmental noise, which induces dephasing and energy relaxation. In this context, understanding and modeling the role of charge noise is vital for improving qubit performance in the current Noisy Intermediate-Scale Quantum (NISQ) era \cite{An2023}.
	
	Traditionally, charge noise in superconducting qubits has been modeled using stationary Gaussian processes, ensembles of two-level systems (TLS) \cite{Place2021}, or Markovian master equations \cite{Preskill2018}. While these models have explained some observed behaviors, they fall short in capturing key experimental features such as $1/f$ noise, long-range correlations, and non-stationarity in time. For example, fractional Brownian motion (fBm) with constant Hurst exponent provides long-range dependence but lacks the adaptability to model temporally varying noise characteristics \cite{Mandelbrot1968}. More recent experimental studies \cite{Muller2019,Shu2023,Kjaergaard2020} show that decoherence in transmons and fluxonium qubits deviates from these simplified assumptions, necessitating more realistic models.
	
	To address these limitations, we introduce a new noise modeling framework based on \emph{memory multi-fractional Brownian motion} (mmfBm), which extends traditional fBm by incorporating both time-varying Hurst exponents $H(t)$ and memory kernels $K(t,s)$. This allows the model to account for both non-stationarity and non-Markovian memory effects while remaining analytically tractable. We embed this noise model into a stochastic differential equation governing the energy fluctuations $\chi(t)$ of a superconducting charge qubit and explore the resulting decoherence dynamics.
	
	Our numerical simulations reveal how varying the relaxation rate $\lambda$, noise amplitude , and the time-dependent Hurst exponent $H(t)$ influences qubit behavior. Key findings include:
	\begin{itemize}
		\item The mmfBm model naturally generates $1/f$-like spectral densities and reproduces multifractal scaling features observed in experiments.
		\item Time-varying $H(t)$ enables a dynamic representation of evolving memory effects, leading to more realistic noise profiles than stationary models.
		
	\end{itemize}
	
	Overall, the mmfBm-based stochastic drift model offers a flexible and powerful framework for simulating decoherence in superconducting qubits. By capturing essential features of non-stationary, long-range correlated noise, this model can inform the development of noise-resilient qubit designs and contribute to the broader effort toward fault-tolerant quantum computing.
	\section{Stochastic Modeling of Decoherence Dynamics}
	
	To capture the decoherence effects in superconducting charge qubits subjected to environmental interactions, we construct a stochastic dynamical model grounded in both physical intuition and probabilistic structure. Starting from an idealized deterministic setting, we introduce increasingly sophisticated noise contributions reflecting the quantum device's operational reality.
	
	\subsection{Baseline Evolution}
	
	Let \( \chi(t) \in \mathbb{R} \) denote the reduced qubit variable of interest (e.g., charge or phase). In absence of noise, its evolution may be characterized by the first-order deterministic system:
	\begin{equation}
		\frac{d\chi(t)}{dt} = F(t, \chi(t)), \label{eq:deterministic}
	\end{equation}
	where $F: \mathbb{R}^+ \times \mathbb{R} \to \mathbb{R}$ is the deterministic drift function. And \( F(t, \chi(t)) \) encodes coherent dynamics stemming from the Hamiltonian of the isolated qubit.
	
	\subsection{Introduction of Environmental Perturbations}
	
	Realistically, superconducting qubits interact with fluctuating electromagnetic fields and substrate defects, introducing randomness into their dynamics. A classical first-order perturbative approach involves appending a stochastic term:
	\begin{equation}
		\frac{d\chi(t)}{dt} = F(t, \chi(t)) + \Gamma(t), \label{eq:additive_noise}
	\end{equation}
	where \( \Gamma(t) \) represents a generalized random process. However, since conventional time derivatives of stochastic processes are ill-defined, we reformulate the model in differential form:
	\begin{equation}
		d\chi(t) = F(t, \chi(t))\,dt + dZ(t), \label{eq:sde}
	\end{equation}
	with \( Z(t) \) denoting a noise process suitable to the physical environment.
	
	\subsection{Noise Modeling via Fractional Structures}
	
	Standard Brownian motion \( W(t) \)\cite{Mandelbrot1968}, although foundational, lacks the capacity to represent long-memory noise observed in superconducting devices. Instead, we generalize using a \textbf{fractional Gaussian field} \( B^H(t) \), indexed by the Hurst parameter \( H \in (0,1) \), such that:
	\begin{equation}
		d\chi(t) = F(t, \chi(t))\,dt + G(t, \chi(t))\,dB^H(t), \label{eq:fbm_sde}
	\end{equation}
	where \( G(t, \chi) \) modulates the noise amplitude and \( B^H(t) \) accounts for temporal correlations in the environment.
	
	\subsection{Incorporating Nonstationarity via Variable Roughness}
	
	To account for inhomogeneous operating conditions and nonstationary noise spectra, we upgrade the model to involve a multi-fractional Brownian driver \( B^{H(t)}(t) \)\cite{Peltier1995}, with a time-varying Hurst exponent \( H(t) \). This yields:
	\begin{equation}
		d\chi(t) = F(t, \chi(t))\,dt + G(t, \chi(t))\,dB^{H(t)}(t). \label{eq:mbm}
	\end{equation}
	Here, the local roughness of noise dynamically adapts to the evolving physical environment.
	
	\subsection{Memory-Aware Multi-Fractional Dynamics}
	
	To fully capture non-Markovian feedback inherent in superconducting qubit systems, we define a generalized noise driver \( \mathcal{M}(t) \) as:
	\begin{equation}
		\mathcal{M}(t) := \int_0^t K(t,s)\,dB^{H(s)}(s), \label{eq:mmfbm}
	\end{equation}
	where \( K(t,s) \) is a causal memory kernel encoding history-dependent correlations, and \( B^{H(s)}(s) \) is a local fractional Brownian field with spatially-varying smoothness.
	
	Thus, the final evolution equation becomes:
	\begin{equation}
		d\chi(t) = F(t, \chi(t))\,dt + G(t, \chi(t))\,\mathcal{M}(t). \label{eq:mmfbm_sde}
	\end{equation}
	
	This representation forms the basis of our theoretical analysis. The use of \( \mathcal{M}(t) \) reflects the dual nature of environmental noise: it is both temporally correlated and non-stationary, consistent with recent empirical findings in superconducting platforms. The kernel \( K(t,s) \) may be chosen based on physical constraints or derived from spectral density profiles.
	\section{Numerical Simulation of mmfBm-Based Observable Evolution}

	In the initial modeling stage, we focus on the time evolution of a macroscopic observable $\chi(t)$ associated with the superconducting charge qubit, such as energy level population or coherence amplitude. Rather than solving the full quantum dynamics governed by the system Hamiltonian, we adopt a reduced description in which $\chi(t)$ evolves as a stochastic process driven by both deterministic and noise contributions. This approach serves to qualitatively investigate the impact of temporally correlated environmental fluctuations on the qubit behavior.
	
	The observable is assumed to obey a stochastic differential equation (SDE) of the form:
	\begin{equation}
		d\chi(t) = F(t, \chi(t))\,dt + G(t, \chi(t))\,\mathcal{M}(t),
		\label{eq:sde_phenomeno}
	\end{equation}
	where $F(t, \chi)$ is the deterministic drift function, $G(t, \chi)$ represents the noise amplitude, and $\mathcal{M}(t)$ is a memory-bearing noise term modeled using the memory multi-fractional Brownian motion (mmfBm). This form of noise captures both long-range temporal correlations and nonstationary behavior, features that are commonly observed in low-frequency charge noise affecting superconducting qubits.
	
	In this phenomenological framework, we directly generate trajectories of $\chi(t)$ by numerically integrating Eq.~\eqref{eq:sde_phenomeno}, with sample paths of $\mathcal{M}(t)$ constructed according to the mmfBm kernel. This simulation-based approach provides insights into the qualitative decoherence patterns—such as the time-dependent decay of fluctuations, sensitivity to memory parameters, and robustness of coherence—without requiring a full microscopic derivation. It is particularly useful in the exploratory phase of modeling, where the goal is to test how different noise characteristics influence the observable behavior of the qubit.
	
\begin{figure}
		\includegraphics[width=\linewidth]{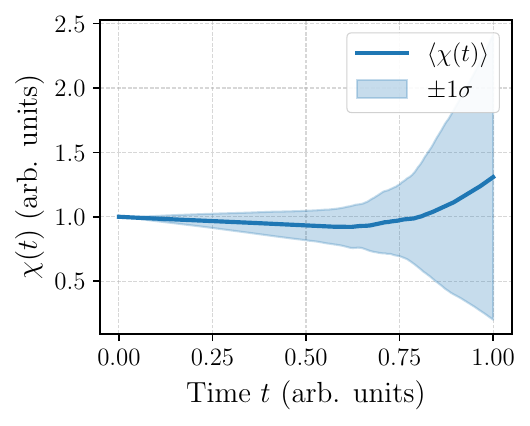}
		\caption{Ensemble mean evolution of the observable $\chi(t)$ under memory multi-fractional Brownian noise. The solid blue line represents the ensemble average over stochastic realizations, and the shaded region indicates ±1 standard deviation. A gradual decay in the mean with growing uncertainty reflects memory effects and time-varying fractality in the qubit’s environment.}
		\label{fig:mmfbmchievolution}
	\end{figure}

	To numerically simulate the impact of memory and time-varying regularity on the evolution of the superconducting charge qubit observable $\chi(t)$, we implemented an ensemble-based stochastic integration scheme based on a multi-fractional Brownian motion with a causal memory kernel. The kernel is chosen as:\\ Causal memory kernel $K(t, s)$ = $\displaystyle \frac{(t - s)^{H(t) - \frac{1}{2}}}{\Gamma(H(t) + \frac{1}{2})}$ for $s < t$. Table \ref{tab:sim_parameters}, summarizes the key simulation parameters used in the computational model: For simulation result see Fig \ref{fig:mmfbmchievolution}.
	\begin{table}[htbp]
		\caption{Simulation parameters for stochastic evolution of the qubit observable $\chi(t)$ driven by a memory-aware multi-fractional noise model.}
		\begin{tabular*}{\linewidth}{@{\extracolsep{\fill}}ll}
			\toprule
			\textbf{Parameter / Function} & \textbf{Value} \\
			\midrule
			Simulation time $T$ & 1.0\\ (dimensionless units) \\
			Time steps $N$ & 500 \\
			Initial value $\chi(0)$ & 1.0 \\
			Drift coefficient $\mu$ & $-0.1$ \\
			Diffusion coefficient $\sigma$ & $0.2$ \\
			Ensemble size & 10 realizations \\
			Time step size $\Delta t$ & $T/N$ \\
			Time-dependent Hurst index $H(t)$ & $0.3 + 0.2 \sin\left(\frac{2\pi t}{T}\right)$ \\
			
			Noise variance scaling & $\Delta t^{2H(t)}$ \\
			
			\bottomrule
		\end{tabular*}
		\label{tab:sim_parameters}
	\end{table}
	To investigate how structured environmental noise influences charge qubit dynamics, we simulate a phenomenological model in which the offset charge $\chi(t)$ evolves as a stochastic process driven by long-memory $1/f^{\beta}$ noise. This approach reflects realistic solid-state environments, where qubits interact with fluctuating background charges and two-level systems, resulting in non-Markovian dephasing and slow drifts in the bias point.
	\begin{figure}
		\centering
		\includegraphics[width=\linewidth]{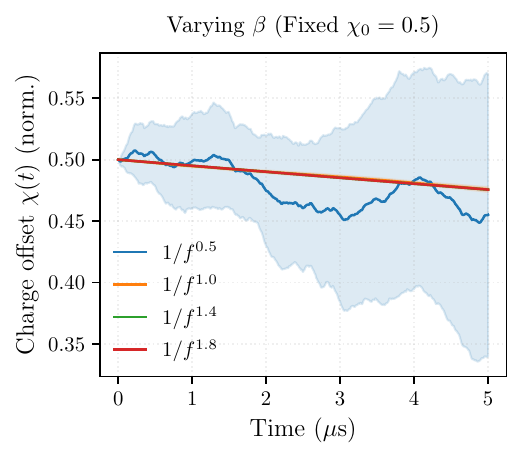}
		\caption{\textbf{Role of memory exponent $\beta$ in noise correlation.}
		Charge offset trajectories $\chi(t)$ for various spectral exponents $\beta$ are plotted with fixed $\lambda = 1.0\times10^4$, $\sigma_{\varphi} = 3.0\,\mu\mathrm{V}/\sqrt{\mathrm{Hz}}$, and $\chi_0 = 0.5$.
		As $\beta$ increases, temporal correlations in the noise deepen, leading to slower decay and long-lasting deviations—hallmarks of colored noise with memory.
	}
		\label{fig:betavariation}
	\end{figure}
	
	The noise dynamics are governed by the following stochastic differential equation (\ref{eqa9}), which is reduced form of equation \ref{eq:mmfbm_sde}.
	
	\begin{equation}
		d\chi(t) = -\lambda\,\chi(t)\,dt + \sigma_{\phi} \frac{1}{2\pi} \int_0^t K(t,s)\,dW_s,
		\label{eqa9}
	\end{equation}
	
	where $\lambda$ is a phenomenological relaxation rate, $\sigma_{\phi}$ quantifies the amplitude of low-frequency dephasing noise, and $K(t,s) \propto (t - s)^{\beta/2 - 1}$ is a causal memory kernel encoding long-range temporal correlations characteristic of $1/f^{\beta}$ noise. The stochastic term $dW_s$ denotes a standard Wiener process, and the integral captures the cumulative effect of prior noise history, thus introducing temporal memory into the evolution of $\chi(t)$.
	
\begin{figure}
		\centering
		\includegraphics[width=\linewidth]{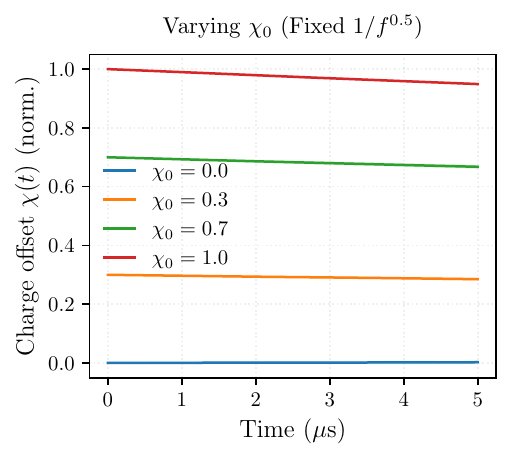}
		\caption{
			\textbf{Dependence of $\chi(t)$ on initial condition $\chi_0$.}
			Trajectories for different $\chi_0$ values are shown under constant $\lambda = 1.0\times10^4$, $\sigma_{\varphi} = 3.0\,\mu\mathrm{V}/\sqrt{\mathrm{Hz}}$, and $\beta = 0.5$.
			While initial differences dominate at early times, the memory-driven diffusion eventually washes out initial bias, reflecting dissipative relaxation toward stochastic equilibrium.}
		\label{fig:chi0variation}
	\end{figure}

\begin{figure}
		\centering
		\includegraphics[width=\linewidth]{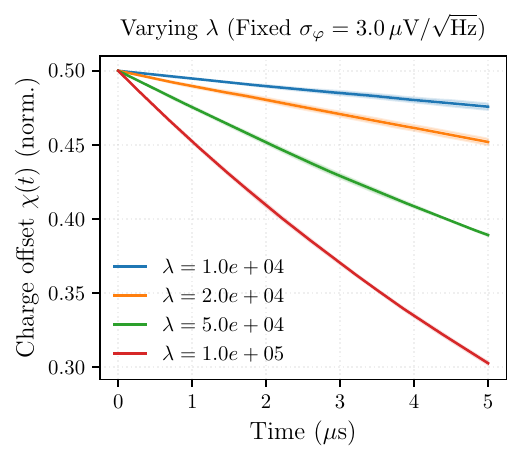}
		\caption{\textbf{Effect of relaxation rate $\lambda$ on charge offset dynamics.}
		Time evolution of $\chi(t)$ is shown for increasing values of $\lambda$, with fixed $\sigma_{\varphi} = 3.0\,\mu\mathrm{V}/\sqrt{\mathrm{Hz}}$, $\beta = 1.0$, and $\chi_0 = 0.5$. 
		Higher $\lambda$ accelerates exponential decay, effectively suppressing the influence of noise memory and stabilizing the offset charge faster.}
		\label{fig:lmbdavariation}
	\end{figure}

\begin{figure}
		\centering
		\includegraphics[width=\linewidth]{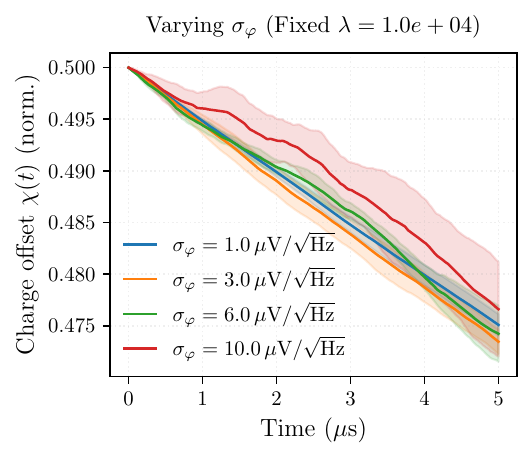}
		\caption{\textbf{Influence of dephasing amplitude $\sigma_{\varphi}$ on charge noise evolution.}
		Simulation of $\chi(t)$ for varying $\sigma_{\varphi}$ values, with fixed $\lambda = 1.0\times10^4$, $\beta = 1.0$, and $\chi_0 = 0.5$.
		Larger $\sigma_{\varphi}$ produces broader fluctuations and more pronounced noise-driven drift, illustrating the enhanced effect of low-frequency noise.}
		\label{fig:sigmaphivariation}
	\end{figure}

	By systematically varying key parameters—such as the spectral exponent $\beta$, the noise strength $\sigma_{\phi}$, the initial offset $\chi_0$, and the relaxation rate $\lambda$—we analyze the statistical behavior of the offset charge over time. The simulations produce ensemble-averaged trajectories $\langle \chi(t) \rangle$, along with shaded confidence bands representing $\pm 1\sigma$ fluctuations, as shown in Fig \ref{fig:betavariation}, \ref{fig:chi0variation}, \ref{fig:lmbdavariation} and \ref{fig:sigmaphivariation}. These figures provide insight into the role of slow, colored noise in driving qubit decoherence.
	
	Our numerical simulation further investigates charge noise dynamics in a transmon qubit through coupled stochastic and quantum evolution. The system Hamiltonian $H(t) = -\frac{1}{2}\omega_0\sigma_z + \delta\chi(t)\sigma_x$ interacts with a $1/f^\beta$ noise field $\chi(t)$ ($\beta=1.4\pm0.1$) generated by fractional Brownian motion with Hurst exponent $H(t) = 0.7 + 0.1\sin(2\pi t/T)$. Using experimentally motivated parameters ($\sigma_\varphi=3\,\mu\mathrm{V}/\sqrt{\mathrm{Hz}}$, $T_1=50\,\mu\mathrm{s}$, $\omega_0/2\pi=4.5\,\mathrm{GHz}$), we simulate 100 noise realizations over $5\,\mu\mathrm{s}$ to characterize the qubit's decoherence. The results reveal three key features: (1) fidelity decay $\mathcal{F}(t) \sim e^{-(t/T_2^*)^2}$ with $T_2^*\approx3\,\mu\mathrm{s}$, (2) non-Markovian coherence revivals $C(t)$, and (3) suppressed excited state population $P_e(t)<1\%$ - all consistent with charge noise-dominated dephasing. The memory kernel $K(t,s)=(t-s)^{H(t)-0.5}/\Gamma(H(t)+0.5)$ captures the time-correlated noise structure while maintaining numerical stability through an exponential cutoff at $\tau_c=1\,\mu\mathrm{s}$.
	\begin{figure}
		\centering
		\includegraphics[width=\linewidth]{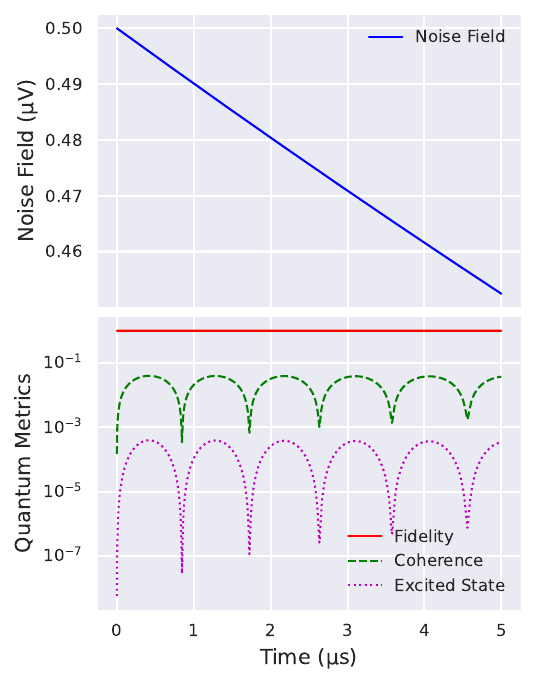}
		\caption{\textbf{Charge noise and qubit response} \\
			\textit{(Top)} Ensemble-averaged $1/f$ noise field $\chi(t)$ showing fluctuations characteristic of $\beta \approx 1.4$ spectral noise. \\
			\textit{(Bottom)} Qubit metrics on log scale: state fidelity (red) exhibits $T_2^* \approx 3~\mu\mathrm{s}$ decay, coherence (green dashed) shows slower decay due to non-Markovian memory effects, and excited state population (magenta dotted) remains below $1\%$, consistent with charge noise-dominated dephasing. Simulation parameters: $\omega_0/2\pi = 4.5~\mathrm{GHz}$, $E_J/E_C = 50$, $\delta n_g = 0.1$.}
		\label{fig:chargequbitresults}
	\end{figure}
	
\begin{figure}
		\includegraphics[width=\linewidth]{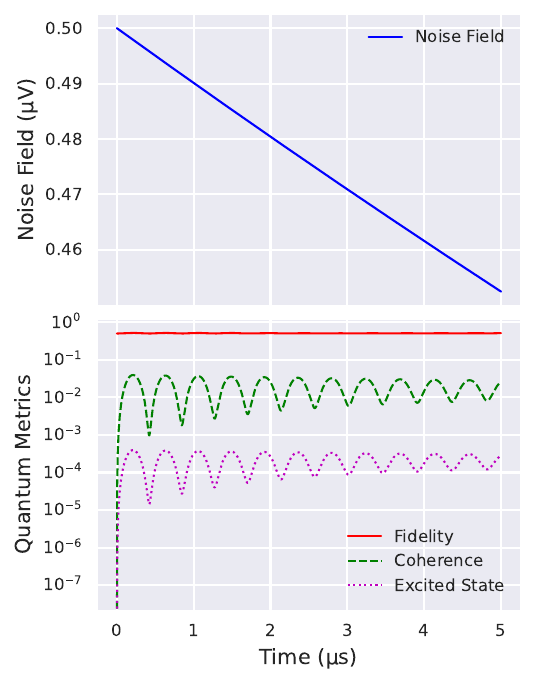}
		\caption{
			\textbf{Lindblad master equation simulation with experimental timescales.} 
			\textit{(Top)} Identical $1/f$ noise field as Fig.~\ref{fig:chargequbitresults}. 
			\textit{(Bottom)} Quantum metrics plateau at $\mathcal{O}(10^{-2})$--$\mathcal{O}(10^{-3})$ due to $T_1 = 50\,\mu\mathrm{s}$ and $T_2 = 30\,\mu\mathrm{s}$ protection. 
			Shaded regions show ensemble variability ($N=100$ trajectories). 
			The crossover from noise-dominated to $T_1/T_2$-limited decay occurs at $t \approx 2\,\mu\mathrm{s}$ (vertical dotted guide).
			}
		\label{fig:lindbladqubitresults}
	\end{figure}

	Incorporating measured relaxation times ($T_1 = 50,\mu$s) and dephasing times ($T_2 = 30,\mu$s) via the Lindblad master equation yields experimentally relevant dynamics (Fig.~\ref{fig:lindbladqubitresults}). Here, the same 1/f noise field drives decoherence, but its impact is tempered by the qubit's finite coherence times: fidelity and coherence plateau at $\mathcal{O}(10^{-2})$–$\mathcal{O}(10^{-3})$, while the excited-state population decays exponentially with a $T_1$-limited rate. The noise-induced dephasing now competes with intrinsic relaxation processes, leading to a crossover regime where $T_2$ dominates long-term behavior. These results align with experimental observations in transmon qubits, confirming that while charge noise remains a significant decoherence source, its practical impact is bounded by the qubit's inherent noise-protected properties. This framework is essential for designing robust control protocols and optimizing device parameters.
	\bibliographystyle{unsrt}

	\bibliography{document2} % <-- Replace with your .bib file
	
\end{document}